\documentclass[10pt,conference]{IEEEtran}
\ifCLASSINFOpdf

\else
\fi
\usepackage{graphicx}
\usepackage[acronym]{glossaries}
\usepackage{cite}
\usepackage{amsmath,amssymb,amsfonts}
\usepackage{algorithmic}
\usepackage{graphicx}
\usepackage{textcomp}
\usepackage{xurl}
\usepackage{multirow}
\usepackage{hyperref}
\usepackage{forest}
\usepackage[table]{xcolor}
\usepackage{tcolorbox}
\usepackage{array}
\usepackage{caption}
\usepackage{changepage}
\usepackage{tabularx}
\usepackage{ragged2e}
\usepackage{url}
\makeglossaries
\usepackage[numbers,sort&compress]{natbib} 

\usepackage{microtype}
\usepackage{xcolor}

\newtcolorbox{findingbox}{
colback=gray!6,
colframe=black!55,
boxrule=0.4pt,
arc=2pt,
left=5pt,
right=5pt,
top=4pt,
bottom=4pt,
fontupper=\small
}

\def\BibTeX{{\rm B\kern-.05em{\sc i\kern-.025em b}\kern-.08em
    T\kern-.1667em\lower.7ex\hbox{E}\kern-.125emX}}
\newacronym{hf}{HF}{Hugging Face}
\newacronym{llm}{LLM}{Large Language Model}
\newacronym{aibom}{AIBOM}{Artificial Intelligence Bill of Materials}
\newacronym{sbom}{SBOM}{Software Bill of Materials}
\newacronym{ml}{ML}{Machine Learning}

\begin{document}
\title{A Large-Scale Measurement of AI Bill of Materials Completeness in Hugging Face Models}

\author{
    Md Erfan, Ahmed Ryan, and Md Rayhanur Rahman\textsuperscript{$\dagger$} \\
    \small Department of Computer Science, The University of Alabama, Tuscaloosa, USA \\
    \small Email: \{merfan, aryan9\}@crimson.ua.edu, mrahman87@ua.edu \\
    \small \textsuperscript{$\dagger$}Corresponding Author: Md Rayhanur Rahman
}
\maketitle
\markboth{Journal of \LaTeX\ Class Files,~Vol.~14, No.~8, August~2015}%
{Shell \MakeLowercase{\textit{et al.}}: Bare Demo of IEEEtran.cls for IEEE Journals}
\author{Md Erfan, Ahmed Ryan and Md Rayhanur Rahman\textsuperscript{$\dagger$}%
\thanks{Md Erfan, Ahmed Ryan, and Md Rayhanur Rahman are with the Department of Computer Science, The University of Alabama, Tuscaloosa, USA. Email: \{merfan, aryan9\}@crimson.ua.edu, mrahman87@ua.edu}%
\thanks{Md Kamal Hossain Chowdhury is with the Alabama Water Institute, The University of Alabama, Tuscaloosa, USA. Email: mhchowdhury@crimson.ua.edu}%
\thanks{\textsuperscript{$\dagger$}Corresponding author.}%
}
\maketitle
\markboth{Journal of \LaTeX\ Class Files,~Vol.~14, No.~8, August~2015}%
{Shell \MakeLowercase{\textit{et al.}}: Bare Demo of IEEEtran.cls for IEEE Journals}


\begin{abstract}
Pretrained machine learning (ML) models help developers build ML-intensive software systems without training models from scratch. However, model repositories often provide incomplete machine-readable documentation about model provenance, licenses, datasets, limitations, and external references, creating transparency and governance gaps across the AI supply chain. Artificial Intelligence Bills of Materials (AIBOMs) address these gaps by documenting AI artifacts, including models, metadata, licenses, datasets, model-card information, and external references. Taking public Hugging Face (HF) model repositories as a case study, this paper empirically investigates AIBOM completeness, defined as the extent to which repositories provide AIBOM-relevant information for machine-readable AI supply-chain documentation. We examine approximately 97.5K AIBOM artifacts to assess the extent to which generated AIBOMs: (i) contain required structural and metadata fields, (ii) represent model identity, license, and external-reference information, (iii) capture model-card documentation such as datasets, limitations, safety-risk assessment, and environmental information, and (iv) vary in documentation coverage across repository and artifact characteristics such as task, license availability, dataset declaration, model family, and paper reference. Results indicate that generated AIBOMs provide complete coverage of required AIBOM structure but limited AI-specific documentation completeness. Required fields are fully represented, but model-card, metadata, responsible-use, environmental, limitation, and meaningful-description fields remain weakly represented or missing across generated artifacts. Our findings motivate improved model-card practices, repository-level traceability, and automated AIBOM validation to advance the generation and adoption of more complete AIBOMs.
\end{abstract}


\IEEEpeerreviewmaketitle

\section{Introduction}

The rapid growth of \glspl{llm} and generative AI is changing how software systems are built, reused, and deployed. Instead of developing every model from scratch, developers increasingly rely on pretrained models shared through public model hubs such as \gls{hf}~\cite{jiang2023empirical,fan2023large}. These repositories now host millions of models across tasks such as text generation, image generation, classification, speech recognition, robotics, and multimodal reasoning. As the number and variety of models continue to grow, AI development increasingly depends on pretrained models that are downloaded, fine-tuned, integrated into applications, and redistributed across downstream pipelines. To use a food-supply-chain metaphor, selecting a pretrained AI model without documentation is like selecting a food product without knowing its ingredients, where it was produced, how it was processed, or whether it is safe and suitable for use.

\begin{figure}[t]
\centering
\includegraphics[width=\columnwidth]{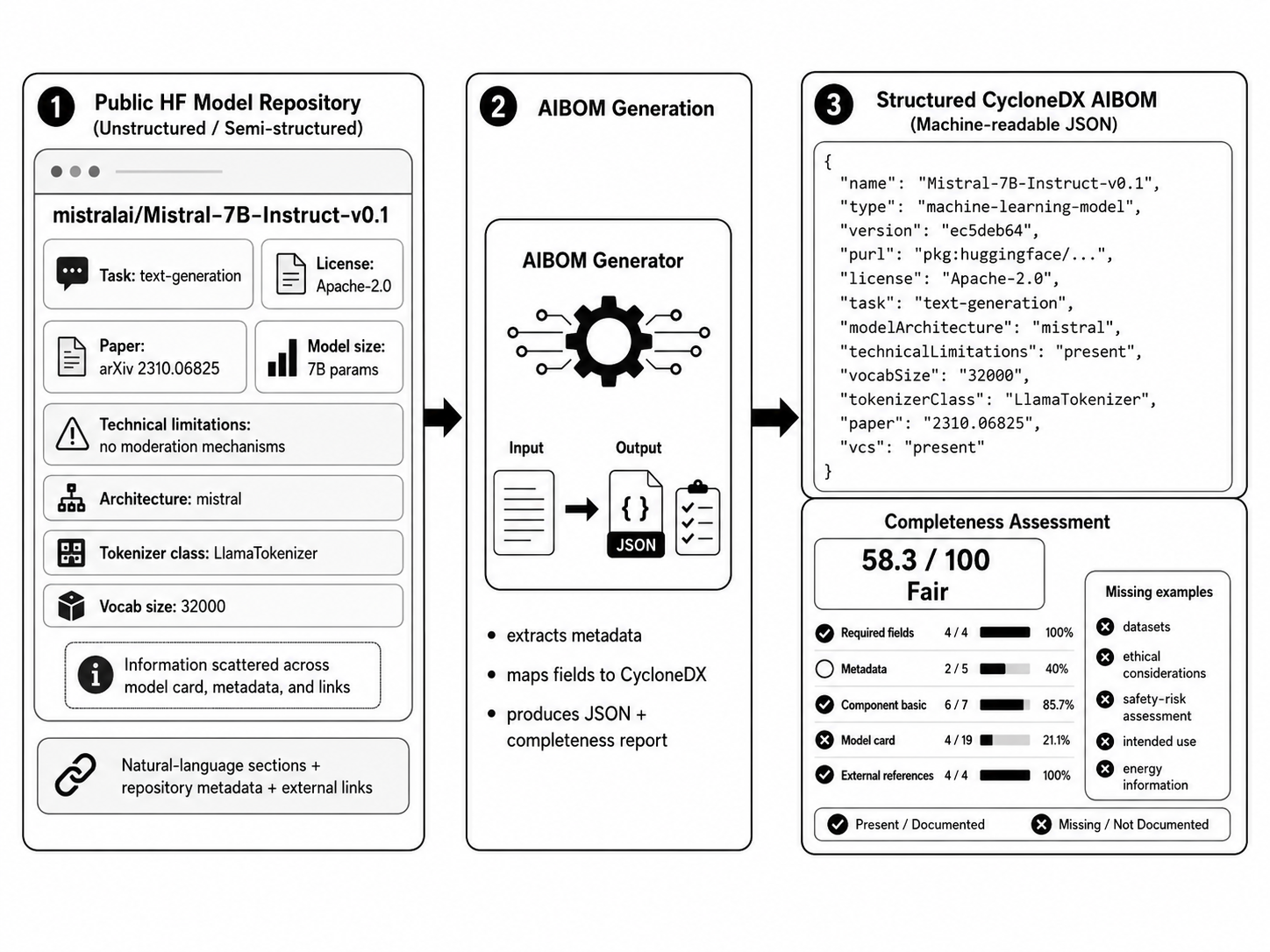}
\caption{Motivating example showing how information from a public \gls{hf} model repository is transformed into a structured AIBOM and evaluated for completeness and readiness.}
\label{fig:motivating-example}
\end{figure}

AI models are reused in high-impact domains such as healthcare, finance, education, scientific research, and public services, where model behavior can affect decisions across downstream software systems. Therefore, downstream users need to know a model’s provenance, training or referenced datasets, license, version, documented limitations, safety risks, and external references. Figure~\ref{fig:motivating-example} illustrates this motivation: information about a public \gls{hf} model may be available, but it is often scattered across model-card text, repository metadata, files, and external links; AIBOM generation organizes this information into machine-readable fields and exposes which transparency fields remain missing. This transparency need has security and governance implications. Prior work has shown that public model repositories can expose users to supply-chain risks, including unsafe model serialization, malicious model files, suspicious configuration files, and repository-level attack vectors~\cite{casey2024large,ryan2025unveiling,jfrog2024malicious,ding2025rusty}. An \gls{aibom} provides machine-readable information for AI model supply-chain transparency at development time~\cite{radanliev2026capability,radanliev2026operationalising}. Without such information, public model reuse becomes difficult to audit, govern, and trust.

Existing research has examined \gls{hf} model reuse~\cite{jiang2023empirical}, repository evolution and maintenance, carbon-footprint reporting~\cite{castano2023exploring}, user perceptions of multimodal LLMs~\cite{liu2026empirical}, model-card documentation~\cite{mitchell2019model,bhat2023aspirations,tsay2020aimmx,crisan2022interactive}, fairness and responsible-AI concerns~\cite{chakraborty2020fairway,wang2026towards,biswas2023fairify}, licensing and model transparency~\cite{pepe2024hugging,german2009license,wu2017analysis}, SBOM adoption and tooling~\cite{hendrick2022software,xia2023empirical,wu2026more,zahan2023software,haque2025security}, and emerging \gls{aibom} frameworks~\cite{bardenstein2023driving,d2025aloha,spoczynski2025atlas,duddu2024laminator,vandendriessche2026aibomgen_arxiv}. These studies provide evidence about how public model repositories grow, how models are documented, and how supply-chain transparency mechanisms are being proposed for AI systems. However, \gls{aibom} completeness, meaning field-level information coverage, remains underexplored at scale. This limitation is consequential because AIBOMs can facilitate governance, traceability, and reuse only when generated artifacts contain sufficient information about models, licenses, datasets, limitations, external references, and AI-specific transparency fields.

\textit{The goal of this study is to support model publishers, AIBOM tool developers, researchers, and end users in advancing AI supply-chain transparency by evaluating AIBOMs generated from Hugging Face models, measuring their completeness and readiness across model repositories, and informing machine-readable AI governance, traceability, and reuse.}

We address the following research questions (RQs).

\textbf{RQ1. How complete are AIBOMs generated from Hugging Face model repositories at scale?}

\textbf{RQ2. Which granular AIBOM fields are most frequently present or missing in generated AIBOM artifacts?}

\textbf{RQ3. How does AIBOM documentation coverage vary across model repository and artifact characteristics?}

We answer these RQs through a large-scale repository-mining study of HF model repositories. We collect a snapshot of 2,942,466 public HF model records and retain 97,940 models with more than 100 downloads for AIBOM generation and analysis. We then generate AIBOM artifacts using the OWASP GenAI Security Project AIBOM Generator and extract overall and category-level completeness scores. Finally, we parse the AIBOM artifacts to measure individual field availability and compare documentation coverage across repository and artifact characteristics, including task, license, dataset declaration, paper reference, model family, architecture, parameter size, and format. This methodology evaluates AIBOM completeness at the score, category, field, and repository-characteristic levels across generated artifacts for downstream transparency.

In summary, our paper makes the following contributions:

\begin{itemize}
\item We generate a large-scale empirical dataset of 97,940 AIBOM artifacts and corresponding completeness score records from \gls{hf} model repositories for downstream transparency analysis and evaluation.

\item We empirically evaluate AIBOM completeness across five major documentation categories: required fields, metadata, component-basic information, model-card documentation, and external traceability references.

\item We analyze whether individual AIBOM fields are present or missing in generated artifacts and identify gaps in AI-specific transparency and readiness fields, including documented limitations, safety-risk assessment, intended use, ethical considerations, environmental information, datasets, licenses, and scholarly paper references.

\item We examine how AIBOM documentation coverage varies across repository and artifact characteristics, including task, license availability, dataset declaration, paper reference, model family, architecture, parameter size, and format, and discuss the implications of these gaps for AI supply-chain transparency, traceability, governance, and reuse. We make experimental data and analysis available at: \textcolor{blue}{https://figshare.com/s/185c21d7bf411fa4b952}

\end{itemize}

The paper is organized as follows. Section~\ref{sec:keyconcepts_related_word} presents key concepts and related work. Section~\ref{sec:methodology} describes the study design, including dataset construction, AIBOM generation, completeness scoring, artifact parsing, and analysis procedures. Section~\ref{sec:results} presents results and key findings. Section~\ref{sec:implications} discusses implications for AIBOM tooling, model reuse, and supply-chain governance. Section~\ref{sec:threats} discusses threats to validity. Section~\ref{sec:conclusion} concludes the paper with future work.

\section{Key Concepts and Related Work}
\label{sec:keyconcepts_related_word}
We provide key concepts with related work discussion in this section to position our study.

\subsection{Key Concepts}
\label{sec:background}

This section provides background on the \gls{hf} Hub, \glspl{aibom}, and the OWASP AIBOM Generator. These concepts frame our study of AIBOM completeness across \gls{hf} model repositories.

\subsubsection{The Hugging Face Hub}
\label{sec:background-hf-hub}

Training and deploying \gls{ml} models requires substantial computational resources, domain knowledge, and engineering effort~\cite{jiang2023empirical, yasmin2026software}. Model reuse therefore plays an important role in modern AI development, because pretrained models can be downloaded, fine-tuned, integrated, and redistributed across downstream software systems~\cite{synovic2026empirical}. The \gls{hf} Hub facilitates model reuse by providing a public platform for hosting, sharing, versioning, and accessing machine learning models and datasets~\cite{huggingface_hub}. Each model repository can include model files, configuration files, tokenizer files, tags, license information, dataset references, evaluation results, external links, and model-card documentation. The repository structure allows model files and documentation artifacts to be versioned and updated over time.

Model cards are central to documentation on the \gls{hf} Hub. A model card can describe model purpose, training data, evaluation metrics, intended use, limitations, and potential biases~\cite{mitchell2019model}. End users often rely on such repository-level documentation to assess whether a model is suitable for reuse, fine-tuning, deployment, or governance review~\cite{donald2026towards}. Model cards and \glspl{aibom} are related, but they serve different purposes. Model cards primarily document model behavior and intended use, whereas an \gls{aibom} organizes AI supply-chain information into a machine-readable format. Therefore, model-card information can provide important input for AIBOM generation, while the resulting \gls{aibom} can support broader transparency, traceability, governance, and reuse.

\subsubsection{Artificial Intelligence Bill of Materials}
\label{sec:background-aibom}

A \gls{sbom} provides a structured inventory of software components, dependencies, licenses, and metadata~\cite{kanemoto2026empirical, wu2026more, wang2026large}. \glspl{sbom} support software supply-chain transparency by helping organizations identify dependencies, evaluate security risks, and respond to security and compliance issues. An \gls{aibom} extends this idea to AI and machine learning systems. AI systems can depend on pretrained models, datasets, framework libraries, configuration files, training environments, evaluation artifacts, and documentation. These assets create a broader supply-chain surface than conventional software dependencies. An \gls{aibom} captures AI-specific assets in a machine-readable format, enabling consistent examination of model provenance, licensing, dataset usage, external references, and transparency information for downstream governance tasks, audits, and reuse decisions~\cite{d2025aloha,cyclonedx_standard,nocera2025we}.

The need for \glspl{aibom} increases as AI models move through reuse, fine-tuning, integration, and redistribution before deployment in software systems. Model users need to know which datasets influenced a model, which license governs reuse, which base model was used, which external resources are linked, and whether limitations or evaluation information are available~\cite{mitchell2019model}. Incomplete documentation can limit security review, compliance assessment, reproducibility analysis, and downstream governance~\cite{tang2026navigating}. A structured \gls{aibom} can reduce this uncertainty by organizing available model information into a machine-readable representation~\cite{radanliev2026operationalising, mortensen2026beyond}.

\subsubsection{OWASP AIBOM Generator}
\label{sec:background-owasp-aibom-generator}

The OWASP GenAI Security AIBOM Generator is an open-source tool for generating \glspl{aibom} for models hosted on \gls{hf}~\cite{owasp_aibom_generator}. The tool accepts a \gls{hf} model identifier as input and extracts available repository information, including model metadata, model descriptions, model-card content, license information, configuration details, and external references when available. The extracted information is organized into a machine-readable \gls{sbom} using the CycloneDX JSON format~\cite{cyclonedx_standard}.

The tool also reports an AIBOM completeness score that measures field-level information coverage across five categories. Required fields capture the minimal CycloneDX structure needed for model identification, while metadata captures information about the \gls{aibom} and model purpose. Component-basic information represents model identity and licensing details. Model-card documentation captures AI-specific transparency information, including model description, parameters, evaluation, limitations, and ethical considerations, while external references capture links to source code, datasets, documentation, model files, and traceability artifacts. Together, these categories produce a completeness score out of 100. Section~\ref{sec:completeness-scoring} describes the weights and scoring procedure.

\subsection{Related Work}

This section positions work in relation to repository mining, model documentation, and AI supply-chain governance.

\subsubsection{Empirical Studies on Hugging Face Model Repositories}

\gls{hf} has become an important empirical setting for studying pretrained ML model reuse, repository evolution, documentation, sustainability, and user-facing challenges. Prior work has examined how practitioners select and reuse pretrained models, which repository attributes support reuse, and what risks arise from incomplete provenance, inconsistent performance claims, and limited documentation~\cite{jiang2023empirical}. Repository-mining studies have further analyzed model-card descriptions, metadata, commit activity, task trends, framework usage, maintenance signals, carbon-reporting practices~\cite{castano2023exploring, chadli2024environmental}, user discussions~\cite{liu2026empirical}, and downstream pre-trained models dependency evolution~\cite{banyongrakkul2026ai}, showing that \gls{hf} functions as an evolving software ecosystem rather than only a model-hosting platform for AI reuse and governance workflows~\cite{jiang2023empirical,castano2023exploring,pepe2024hugging,liu2026empirical, castano2024lessons}.

Other empirical work has focused on specific transparency concerns in the \gls{hf} ecosystem. Castaño et al.~\cite{castano2023exploring} studied carbon-emission reporting and carbon-efficiency factors, while Liu et al.~\cite{liu2026empirical} analyzed user discussions around general-purpose and multimodal LLMs, identifying concerns such as access barriers, generation quality, deployment complexity, documentation limitations, and resource constraints. These studies characterize model reuse, maintenance, sustainability, and user-facing concerns. Our work builds on this repository-mining perspective but shifts the analysis target to whether public \gls{hf} repositories contain sufficient structured information to support complete CycloneDX-based AIBOM generation.

\subsubsection{ML Model Documentation, Transparency, and Responsible AI}

Model documentation has been widely studied as a mechanism for improving transparency, accountability, and responsible reuse of ML models. Mitchell et al.~\cite{mitchell2019model} introduced model cards as structured documentation for reporting model details, intended uses, evaluation conditions, and limitations. Later work examined how model documentation is produced, maintained, extracted, and presented in practice across repository settings. Bhat et al.~\cite{bhat2023aspirations} analyzed model descriptions from Hugging Face, GitHub, and industrial sources, and proposed DocML to guide model-card creation and maintenance. Tsay et al.~\cite{tsay2020aimmx} proposed AIMMX for extracting AI model metadata from repositories. Crisan et al.~\cite{crisan2022interactive} studied interactive model cards and examined how interactive documentation affects understandability, interpretability, and trust. These studies focus on improving, extracting, or presenting model documentation, while our study examines how available documentation is represented in generated CycloneDX AIBOM artifacts.

Transparency studies on \gls{hf} have examined dataset, bias, and license documentation. Pepe et al.~\cite{pepe2024hugging} analyzed how \gls{hf} models document training datasets, fairness limitations, and licenses, showing limited exposure of training datasets, bias information, and licensing details. Their work motivates automated support for AI model transparency and AIBOM generation. Our study extends this line of work by measuring AIBOM completeness across 97,940 AIBOM artifacts.

Responsible-AI research further motivates the need for fields related to bias, safety, limitations, ethical considerations, and intended use. Prior work on fairness and bias mitigation has proposed methods such as fairness-aware re-ranking, fair ML construction, counterfactual bias analysis, and neural-network fairness verification~\cite{chakraborty2020fairway,wang2026towards,biswas2023fairify}. These techniques address fairness at the algorithmic, testing, or verification level. Our work does not propose a fairness-mitigation method; instead, our work examines whether generated AIBOMs contain the structured documentation needed to make responsible-use and safety-related information visible to downstream users.
\subsubsection{SBOMs, AIBOMs, and Software Supply-Chain Transparency}

\glspl{sbom} provide structured inventories of software components, dependencies, licenses, and related metadata. SBOM research has examined cybersecurity readiness, practitioner adoption, SBOM completeness, tool limitations, and ecosystem readiness~\cite{hendrick2022software,xia2023empirical,wu2026more,zahan2023software}. These studies show that SBOM generation and consumption involve content, tooling, maintenance, interoperability, and adoption issues. AI systems introduce additional requirements because models depend on datasets, training procedures, model cards, hyperparameters, evaluation information, and external resources.

AIBOMs extend SBOM concepts to AI and ML systems. CycloneDX and SPDX have begun incorporating AI-specific metadata, while AIBOM-related work emphasizes model identity, datasets, licenses, dependencies, intended use, ethical considerations, and provenance information. AIBOMs are therefore relevant for AI supply-chain transparency, compliance review, reuse, and governance. Our study focuses on the empirical question of how complete generated CycloneDX-based AIBOMs are when created from Hugging Face repositories.

Tool-based AIBOM work is related to our study. ALOHA generates AIBOMs for \gls{hf} models by parsing model-card metadata and mapping extracted information to CycloneDX fields~\cite{d2025aloha}. ALOHA evaluates AIBOMs on a 312-model sample and reports limitations related to metadata completeness and model-card standardization. Our study follows the same motivation, but expands the scale and analyzes completeness across 97,940 generated AIBOM artifacts. Other AIBOM frameworks emphasize verifiability, provenance, and attestation. AIBoMGen generates signed AIBOMs during model training and captures datasets, model metadata, environment details, hashes, signatures, and in-toto attestations~\cite{vandendriessche2026aibomgen_arxiv}. Atlas captures ML lifecycle provenance using trusted execution environments and transparency logs~\cite{spoczynski2025atlas}. Laminator generates verifiable ML property cards through hardware-assisted attestations~\cite{duddu2024laminator}. AICert binds training configuration, dataset hashes, and model outputs into hardware-rooted proof artifacts. These frameworks strengthen provenance and integrity guarantees. Our study addresses a complementary repository-level question: whether public model repositories provide sufficiently complete information for standardized AIBOM artifacts.

\subsubsection{Licensing, Provenance, and Reuse of Open-Source AI Artifacts}

Licensing and provenance have long been studied in open-source software. Prior work has examined license adoption, license compatibility, license-integration patterns, license inconsistency, license recommendation, and licensing bugs in software ecosystems~\cite{german2009license,wu2017analysis}. These studies show that reuse decisions depend not only on component availability, but also on license clarity, compatibility, and provenance. AI model reuse raises related concerns because pretrained models can carry model-specific licenses, dataset restrictions, responsible-use terms, and downstream redistribution constraints.

\gls{hf} transparency studies extend these concerns to AI artifacts. Pepe et al.~\cite{pepe2024hugging} studied dataset declarations, bias documentation, model licenses, and possible incompatibilities between \gls{hf} models and GitHub client projects, showing that incomplete licensing information or dataset provenance can create legal and transparency risks. Provenance also connects licensing, dataset declarations, paper references, repositories, and version-control links, which downstream users need for reuse, compliance, traceability, and auditability~\cite{hohensinner2026tracing,radanliev2026operationalising,vandendriessche2026aibomgen_arxiv}. AIBOMs provide a machine-readable representation for these signals, but their quality depends on the information available in source repositories. Our RQ3, therefore, complements prior licensing and provenance work by analyzing how AIBOM documentation coverage varies across license, dataset declaration, paper reference, task, and traceability signals.

\section{Methodology}
\label{sec:methodology}
This section describes our methodology, including \gls{hf} metadata collection, model filtering, AIBOM generation, completeness scoring, artifact parsing, and data analysis.
\subsection{Hugging Face Metadata Collection}
\label{sec:hf-metadata-collection}

We implemented a Python-based data collection pipeline to retrieve metadata from \gls{hf}, a widely used model hub for pretrained \gls{ml} and generative AI models. The pipeline uses the \gls{hf} Hub API to collect model-level information available from model repositories. For each model, we collected metadata such as the model identifier, author, task, library name, number of downloads, likes, creation date, last modification date, commit SHA, license, languages, datasets, evaluation metrics, base model information, model type, and architecture information.

Initially, we collected all available \gls{hf} model entries returned by the API, regardless of whether they contained metadata or model-card documentation. This allowed us to first characterize the broader \gls{hf} model ecosystem and then apply filtering criteria for the AIBOM analysis. Since AIBOM generation is intended to assess documentation completeness, we did not manually fill missing fields in the model repositories. Instead, we preserved missing, incomplete, or inconsistent metadata as part of the empirical signal, because such gaps directly affect \gls{aibom} completeness and AI supply-chain transparency.

\subsection{Data Filtering and Preprocessing}
\label{sec:filtering-preprocessing}
Generating AIBOMs for all \gls{hf} repositories is computationally expensive and constrained by API rate limits. Therefore, we retained models with more than 100 downloads for the AIBOM generation stage. This threshold focuses the analysis on repositories with observable reuse while keeping the study computationally feasible and preserving a large sample across tasks, libraries, licenses, and model families. Models with very low or no downloads were excluded because they provide limited evidence of reuse and would substantially increase the cost of large-scale AIBOM generation.

We organized the filtered dataset into multiple CSV files, each containing approximately 10,000 model entries, to support scalable and incremental AIBOM generation. This batching strategy allowed us to run the AIBOM generation pipeline incrementally and reduced the risk of losing progress because of API rate limits, network interruptions, or tool execution failures. For model-size information, we extracted parameter counts where available, using metadata derived from safetensors files when such information was present in the repository. 

\subsection{AIBOM Generation and Validation}
\label{sec:aibom-generation}

For all filtered models, we generated an \gls{aibom} using the OWASP GenAI Security Project AIBOM Generator~\cite{owasp_aibom_generator}. The tool accepts a \gls{hf} model identifier or model URL as input. The generator then accesses publicly available repository information, including model metadata, model-card content, license information, configuration details, and external references when available. The extracted information is organized into a CycloneDX-based \gls{aibom} in machine-readable JSON format~\cite{cyclonedx_standard}. In this format, the \gls{aibom} represents the model as a component and records associated metadata, licenses, documentation fields, and external references.

We validated each \gls{aibom} against the CycloneDX JSON structure to ensure that only parseable and analyzable artifacts were used for completeness analysis. This validation checked whether each artifact could be parsed as valid JSON and whether it contained expected CycloneDX-level fields, such as \texttt{bomFormat}, \texttt{specVersion}, \texttt{metadata}, and \texttt{components}. Artifacts that failed parsing or lacked score information were excluded from the completeness dataset.

\subsection{AIBOM Completeness Score Calculation}
\label{sec:completeness-scoring}
We used the completeness scoring method from the OWASP AIBOM Generator to measure AIBOM completeness for each HF repository. We did not define a new completeness metric. As described in Section II-A 3, the generator computes a score out of 100 across five documentation categories: required fields, metadata, component-basic information, model-card documentation, and external references. Required fields, metadata, and component-basic information each contribute 20 points; model-card documentation contributes 30 points; and external references contribute 10 points.

Following the OWASP AIBOM Generator scoring method~\cite{owasp_aibom_generator}, we compute each category score, subtotal, and final score as follows:
\begin{equation}
\begin{aligned}
S_i &= \frac{F_{\text{present}, i}}{F_{\text{total}, i}} \times W_i, \\
\text{Subtotal} &= \sum_{i=1}^{5} S_i, \\
\text{FinalScore} &= \text{Subtotal} \times P
\end{aligned}
\label{eq:aibom-score}
\end{equation}
where $S_i$ represents the score for category $i$, $F_{\text{present}, i}$ and $F_{\text{total}, i}$ denote the number of present fields and total fields in category $i$ respectively, $W_i$ is the specific category weight, and $P$ represents the penalty factor. 

The penalty factor accounts for missing critical or important fields. Under the tool-defined scoring method, two to three missing critical fields receive a 10\% penalty, four or more receive a 20\% penalty, and five or more missing important fields receive a 5\% penalty. These penalties are cumulative when both critical and important field penalties apply.

\subsection{Score Extraction and Artifact Parsing}
\label{sec:score-extraction-artifact-parsing}

We developed a Python-based pipeline to integrate AIBOM generation, score extraction, and artifact parsing into our analysis workflow. The pipeline reads the filtered \gls{hf} model identifiers from CSV files and invokes the OWASP GenAI Security Project AIBOM Generator for each model through its command-line interface. For each model identifier, the pipeline specifies an output path for the generated AIBOM artifact. This process links each model to both a machine-readable \gls{aibom} file and a corresponding completeness assessment.

The pipeline was designed for scalable and recoverable execution across long-running batch jobs. The pipeline divides the input models into multiple shards, with each shard representing a smaller partition of the dataset processed independently. Each shard writes its own result file and stores generated AIBOM artifacts in a shard-specific output directory. The pipeline also saves progress after each processed model, allowing execution to resume without repeating completed models.

For each generated \gls{aibom}, the pipeline extracts the overall completeness score and the category-level scores reported by the generator. The extracted categories include \textit{required fields}, \textit{metadata}, \textit{component-basic information}, \textit{model-card documentation}, and \textit{external references}. For each category, the pipeline records the achieved score, total possible score, and percentage score. The pipeline also records the completeness level, AIBOM profile, tool-reported license, generation status, return code, and error message when generation fails.

To check the consistency of the automated score extraction, we inspected a sample of 100 generated AIBOM artifacts and compared the extracted score records with the corresponding score reports produced by the generator. This validation checked whether the overall completeness score, category-level scores, field checklist status, and generated artifact path were consistently recorded in our dataset. The manual check helped confirm that the automated extraction pipeline preserved the score information reported by the generator.

\subsection{Analysis Procedure}
\label{sec:data-analysis}
We analyzed the AIBOM artifacts and score records in relation to the three research questions. The analysis proceeds from overall completeness, to individual field availability, to documentation coverage across repository characteristics.

For RQ1, we analyzed overall and category-level AIBOM completeness using the AIBOM score dataset, which contains 97,940 score records. Each row corresponds to one generated \gls{aibom} and includes the completeness score, completeness level, and category-level scores reported by the generator. For each category, we used three score columns: achieved score, total possible score, and percentage score. We computed descriptive statistics for the completeness score and for each category-level percentage, including count, mean, median, minimum, maximum, standard deviation, first quartile, and third quartile. We also computed Pearson correlations~\cite{benesty2009pearson} between each category-level percentage and the completeness score. Required fields were excluded from correlation analysis because the category remained constant at 100\% across valid score records. These correlations are interpreted as contribution-oriented associations rather than causal relationships because the final score is computed from weighted category scores.

For RQ2, we analyzed granular field-level availability in the parsed AIBOM artifacts. This analysis measured whether required fields, metadata fields, component-basic fields, model-card documentation fields, and external-reference fields were present in each generated \gls{aibom}. We also checked whether model descriptions contained substantive content rather than placeholder values. This field-level analysis explains which specific AIBOM fields are commonly represented, rarely represented, or absent across generated artifacts.

For RQ3, we analyzed how AIBOM documentation coverage varies across repository characteristics. Because RQ3 covers documentation signals, we divided the analysis into four subquestions. RQ3.1 asks whether models with paper references and dataset declarations show different field coverage than models without these signals. RQ3.2 asks how AIBOM documentation coverage varies across model tasks. RQ3.3 asks how combined license--dataset--paper availability is associated with stronger or weaker AIBOM documentation coverage. RQ3.4 asks which model families and artifact characteristics appear among the highest-scoring AIBOMs. For each subquestion, we compared the coverage of AIBOM fields related to traceability, limitations, safety-risk assessment, environmental documentation, dataset disclosure, license information, version-control references, and paper references. We interpret these comparisons as associations between repository documentation practices and AIBOM coverage, rather than as causal effects.

\begin{table}[!t]
\centering
\caption{Score summary for generated \gls{hf} model AIBOMs. All metrics are computed over 97,940 AIBOM artifacts.}
\label{tab:aibom-score-summary}
\footnotesize
\setlength{\tabcolsep}{1.85pt} 
\renewcommand{\arraystretch}{1.1}
\begin{tabular}{l rrrrrrr}
\hline
\textbf{Metric} & \textbf{Mean} & \textbf{Median} & \textbf{Min} & \textbf{Max} & \textbf{Std.} & \textbf{$Q_1$} & \textbf{$Q_3$} \\
\hline
Completeness Score & 54.31 & 54.10 & 46.60 & 68.70 & 4.44 & 50.90 & 58.10 \\
Required Fields (\%) & 100.00 & 100.00 & 100.00 & 100.00 & 0.00 & 100.00 & 100.00 \\
Metadata (\%) & 40.00 & 40.00 & 40.00 & 60.00 & 0.13 & 40.00 & 40.00 \\
Component Basic (\%) & 81.75 & 85.50 & 71.50 & 85.50 & 6.20 & 71.50 & 85.50 \\
Model Card (\%) & 19.51 & 16.67 & 6.00 & 50.00 & 10.48 & 11.00 & 31.67 \\
External Refs (\%) & 68.02 & 75.00 & 50.00 & 100.00 & 17.11 & 50.00 & 75.00 \\
\hline
\end{tabular}
\smallskip
\end{table}
\section{Results}
\label{sec:results}
This section reports the results by first summarizing the scale and composition of the \gls{hf} metadata snapshot and then answering the three research questions.

We begin with the collected HF metadata snapshot. The snapshot contains 2,942,466 public model records, with repository creation increasing sharply from 100,576 models in 2022 to 1,145,805 in 2025; the 2026 count is 632,358 because the snapshot covers only part of the year. The ecosystem is dominated by \texttt{transformers}, \texttt{safetensors}, \texttt{peft}, and \texttt{diffusers}, while the largest task groups are \texttt{text-generation}, \texttt{text-classification}, \texttt{text-to-image}, \texttt{reinforcement-learning}, and \texttt{automatic-speech-recognition}. These descriptive results show that \gls{hf} is a large and rapidly expanding model-reuse platform, motivating our subsequent AIBOM completeness analysis for models with observable reuse.

\subsection{RQ1. How complete are AIBOMs generated from Hugging Face model repositories at scale?}
\label{sec:rq1-results}

\subsubsection{Overall completeness-score distribution}

Table~\ref{tab:aibom-score-summary} summarizes the completeness-score distribution across 97,940 valid AIBOM score records. The generated AIBOMs achieved a mean completeness score of 54.31 out of 100, with a median of 54.10. The first and third quartiles are 50.90 and 58.10, respectively, and the standard deviation is 4.44. The observed scores range from 46.60 to 68.70. This narrow distribution indicates that most generated AIBOMs are moderately complete but do not reach high-completeness levels.

\begin{findingbox}
\textbf{Finding 1.} \textit{Generated AIBOMs are moderately complete overall. The mean score is 54.31 out of 100, indicating that public \gls{hf} model repositories can be converted into AIBOM artifacts at scale, but the resulting artifacts remain far from complete.}
\end{findingbox}

\subsubsection{Category-level completeness}

The category-level results show substantial variation across the five OWASP AIBOM completeness categories. Required fields achieve complete coverage, with a mean and median of 100.00\% across valid records. Component-basic information is also comparatively strong, with a mean of 81.75\% and a median of 85.50\%. External references show moderate but uneven coverage, with a mean of 68.02\% and a standard deviation of 17.11.

In contrast, metadata and model-card documentation are much weaker. Metadata has a mean of 40.00\%, with almost no variation across records. Model-card documentation has the lowest coverage, with a mean of 19.51\% and a median of 16.67\%. This result indicates that generated AIBOMs usually contain the basic structure needed for identification, but they lack richer AI-specific documentation for transparency.

\begin{findingbox}
\textbf{Finding 2.} \textit{AIBOM completeness is uneven across categories. Required fields and component-basic information are well represented, while metadata and model-card documentation are the main sources of incompleteness in generated artifacts.}
\end{findingbox}

\subsubsection{Categories associated with overall completeness}

Table~\ref{tab:category-overall-correlation} shows the correlation between category-level completeness and the AIBOM completeness score. Model-card documentation has the strongest association with overall completeness, with a Pearson correlation of 0.888. External references show a strong association at 0.662. Component-basic information has a weaker association at 0.351, while metadata has no association at 0.003. Required fields are excluded because they are constant at 100.00\% across valid records.

\begin{table}[t]
\centering
\caption{Correlation between AIBOM category-level scores and overall completeness across generated artifacts.}
\label{tab:category-overall-correlation}
\small
\begin{tabular}{lr}
\hline
\textbf{Category} & \textbf{Correlation} \\
\hline
Model-card documentation & 0.888 \\
External references & 0.662 \\
Component-basic information & 0.351 \\
Metadata & 0.003 \\
\hline
\end{tabular}
\end{table}

These correlations indicate that variation in the final score is driven by model-card documentation and external-reference coverage. Metadata contributes little to score variation because most records have nearly identical metadata coverage.

\begin{findingbox}
\textbf{Finding 3.} \textit{Overall AIBOM completeness is driven primarily by model-card documentation and external references. Improving these categories is likely to produce the largest increase in final completeness scores across the analyzed models.}
\end{findingbox}

\subsection{RQ2. Which granular AIBOM fields are most frequently present or missing in generated AIBOM artifacts?}
\label{sec:rq2-results}
\begin{table}[t]
\centering
\caption{AIBOM Field Availability across Artifacts.}
\label{tab:granular-field-availability}
\footnotesize 
\renewcommand{\arraystretch}{1.05} 
\begin{tabularx}{\columnwidth}{l >{\RaggedRight\arraybackslash}X r r}
\hline
\textbf{Category} & \textbf{Field} & \textbf{Present} & \textbf{Present (\%)} \\
\hline
Required Fields & bomFormat & 97,940 & 100.00 \\
Required Fields & specVersion & 97,940 & 100.00 \\
Required Fields & serialNumber & 97,940 & 100.00 \\
Required Fields & version & 97,940 & 100.00 \\
\hline
Metadata & primaryPurpose & 97,940 & 99.99 \\
Metadata & suppliedBy & 97,940 & 99.99 \\
Metadata & standardCompliance & 0 & 0.00 \\
Metadata & domain & 4 & $<0.01$ \\
Metadata & autonomyType & 0 & 0.00 \\
\hline
Component Basic & name & 97,940 & 100.00 \\
Component Basic & type & 97,940 & 100.00 \\
Component Basic & version & 97,940 & 100.00 \\
Component Basic & purl & 97,940 & 100.00 \\
Component Basic & description & 97,940 & 99.99 \\
Component Basic & licenses & 71,640 & 73.14 \\
Diagnostic & meaningfulDescription & 211 & 0.22 \\
\hline
Model Card & datasets & 38,542 & 39.35 \\
Model Card & hyperparameter & 24,884 & 25.40 \\
Model Card & technicalLimitations & 16,393 & 16.74 \\
Model Card & safetyRiskAssessment & 9,555 & 9.76 \\
Model Card & ethicalConsiderations & 0 & 0.00 \\
Model Card & intendedUse & 0 & 0.00 \\
Model Card & modelExplainability & 0 & 0.00 \\
Model Card & informationAboutTraining & 0 & 0.00 \\
Model Card & informationAboutApplication & 0 & 0.00 \\
Model Card & metric & 0 & 0.00 \\
Model Card & SensitivePersonalInformation & 0 & 0.00 \\
Model Card & vocab\_size & 57,785 & 58.99 \\
Model Card & tokenizer\_class & 60,382 & 61.65 \\
\hline
Environmental & energyConsumption & 19 & 0.02 \\
Environmental & energyQuantity & 0 & 0.00 \\
Environmental & energyUnit & 0 & 0.00 \\
\hline
External Refs & paper & 13,228 & 13.50 \\
External Refs & vcs & 57,496 & 58.70 \\
External Refs & website & 97,940 & 99.99 \\
External Refs & downloadLocation & 97,940 & 99.99 \\
\hline
\end{tabularx}
\end{table}

We parsed 97,940 generated AIBOM artifacts and measured the presence of individual fields in order to explain the category-level results. Table~\ref{tab:granular-field-availability} reports field-level results.

\subsubsection{Structural and component-identification fields}

Required CycloneDX fields are fully represented in the parsed artifacts. The fields \texttt{bomFormat}, \texttt{specVersion}, \texttt{serialNumber}, and \texttt{version} appear in all 97,940 artifacts. Component-identification fields are also highly complete: \texttt{name}, \texttt{type}, \texttt{version}, and \texttt{purl} appear in all parsed artifacts. These results explain why required fields and component-basic information receive comparatively strong category-level scores.

\begin{findingbox}
\textbf{Finding 4.} \textit{Generated AIBOMs are structurally complete at the CycloneDX level. Required fields and core model-identification fields are almost universally represented.}
\end{findingbox}

\subsubsection{Metadata, licensing, and description fields}

Basic metadata fields are common, but governance-oriented metadata is largely missing in the generated AIBOMs. The fields \texttt{primaryPurpose} and \texttt{suppliedBy} are present in 99.99\% of artifacts. However, \texttt{standardCompliance} and \texttt{autonomyType} are absent from all parsed artifacts, and \texttt{domain} appears in only four artifacts. Licensing information is present in 71,640 artifacts, corresponding to 73.14\%.

The description field requires separate interpretation in AIBOM completeness analysis. Although \texttt{description} is present in 99.99\% of artifacts, only 211 artifacts contain a meaningful description beyond placeholder content. This corresponds to only 0.22\% of parsed artifacts and shows that field presence can overestimate documentation quality.

\begin{findingbox}
\textbf{Finding 5.} \textit{Basic metadata and descriptions are often syntactically present, but governance metadata and meaningful descriptions are largely missing.}
\end{findingbox}

\subsubsection{Model-card and AI-specific transparency fields}

Model-card fields show the largest documentation gaps. \textit{Dataset information} appears in 38,542 artifacts, corresponding to 39.35\%. \textit{Hyperparameter information} appears in 25.40\%, \textit{technical limitations} in 16.74\%, and \textit{safety-risk assessment} in 9.76\%. Responsible-use and explanation fields are absent: \textit{ethicalConsiderations}, \textit{intendedUse}, \textit{modelExplainability}, \textit{informationAboutTraining}, \textit{informationAboutApplication}, \textit{metric}, and \textit{useSensitivePersonalInformation} appear in 0.00\% of artifacts. Environmental documentation is also nearly absent: \textit{energyConsumption} appears in only 19 artifacts, while \textit{energyQuantity} and \textit{energyUnit} are absent.

\begin{findingbox}
\textbf{Finding 6.} \textit{AI-specific transparency fields are weakly represented or absent. Dataset declarations, limitations, safety information, responsible-use fields, and environmental fields remain major documentation gaps in generated AIBOMs.}
\end{findingbox}

\subsubsection{External-reference and traceability fields}

External references are unevenly represented. Repository-oriented links are nearly complete: \texttt{website} and \texttt{downloadLocation} appear in 99.99\% of artifacts. However, richer traceability fields are less consistent. Version-control references appear in 57,496 artifacts, corresponding to 58.70\%, and paper references appear in 13,228 artifacts, corresponding to 13.50\%. This indicates that generated AIBOMs usually provide links back to the model repository for downstream users, but fewer artifacts connect models to source-code repositories or research papers.

\begin{findingbox}
\textbf{Finding 7.} \textit{External references support basic repository traceability, but traceability through version-control links and especially paper references remains uneven across all artifacts.}
\end{findingbox}

\subsection{RQ3: How does AIBOM documentation coverage vary across model repository and artifact characteristics?}
\label{sec:rq3-results}

RQ3 examines whether AIBOM documentation coverage varies across repository-level and artifact-level characteristics in the analyzed dataset. We focus on paper reference, dataset declaration, task, license availability, model family, architecture, parameter size, and format because these characteristics are directly connected to traceability, reuse, governance, top-score patterns, and model readiness. Rather than reporting each field independently, we use hierarchical audit views and top-score categorization to show how documentation signals and artifact characteristics combine across repository groups.

\subsubsection{RQ3.1: Paper and dataset availability}

\begin{figure}[t]
\centering
\begin{forest}
for tree={
    grow'=0,
    parent anchor=east,
    child anchor=west,
    anchor=west,
    edge path={
        \noexpand\path [draw, \forestoption{edge}]
        (!u.parent anchor) -- +(6pt,0) |- (.child anchor)\forestoption{edge label};
    },
    l sep=12pt,        
    s sep=2pt,         
    inner sep=2pt,
    rounded corners,
    draw,
    align=left,
    font=\scriptsize,  
}
[Total models\\\textbf{97,940}
    [Has paper ref.\\\textbf{13,228} (13.50\%)
        [Has dataset\\\textbf{8,589} (64.93\%)
            [Has license\\\textbf{6,373} (74.20\%)]
            [Has VCS ref.\\\textbf{8,554} (99.59\%)]
            [Has tech limitations\\\textbf{4,212} (49.04\%)]
        ]
        [No dataset\\\textbf{4,639} (35.07\%)
            [Has license\\\textbf{4,018} (86.61\%)]
            [Has VCS ref.\\\textbf{4,601} (99.18\%)]
            [Has tech limitations\\\textbf{952} (20.52\%)]
        ]
    ]
    [No paper ref.\\\textbf{84,722} (86.50\%)
        [Has dataset\\\textbf{29,953} (35.35\%)
            [Has license\\\textbf{25,147} (83.95\%)]
            [Has VCS ref.\\\textbf{14,194} (47.39\%)]
            [Has tech limitations\\\textbf{7,309} (24.40\%)]
        ]
        [No dataset\\\textbf{54,769} (64.65\%)
            [Has license\\\textbf{36,102} (65.92\%)]
            [Has VCS ref.\\\textbf{30,147} (55.04\%)]
            [Has tech limitations\\\textbf{3,920} (7.16\%)]
        ]
    ]
]
\end{forest}
\caption{Audit-style tree of AIBOM documentation coverage by paper-reference and dataset availability.}
\label{fig:aibom-audit-tree-paper-dataset}
\end{figure}
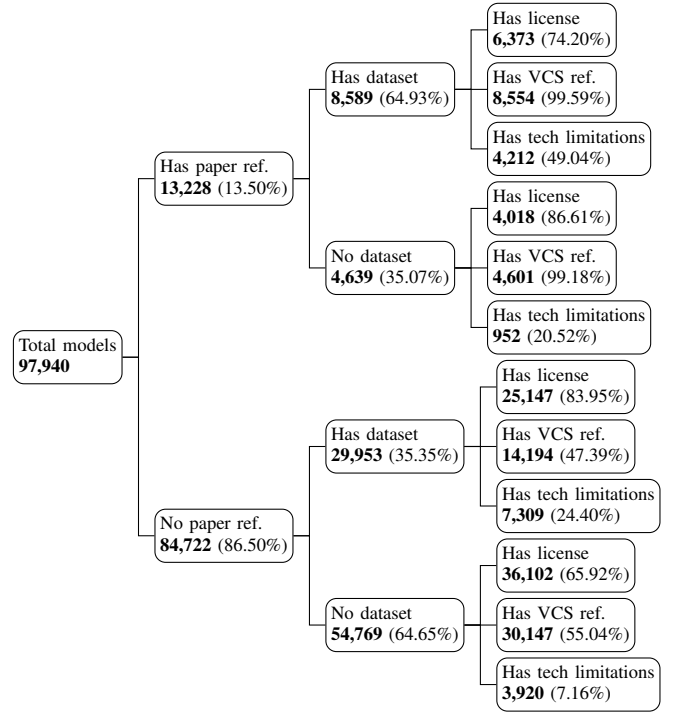

Figure~\ref{fig:aibom-audit-tree-paper-dataset} presents an audit-style tree of AIBOM documentation coverage by paper-reference and dataset availability. Paper references are present in 13,228 artifacts, representing 13.50\% of the 97,940 parsed AIBOM artifacts. Among models with paper references, 64.93\% also include dataset information. In contrast, among models without paper references, only 35.35\% include dataset information. This pattern indicates that paper-linked models are more likely to provide dataset declarations.

The same pattern appears for traceability and limitation fields. Among models with both paper references and dataset declarations, 99.59\% include VCS references and 49.04\% include technical limitations. Among models without paper references and without dataset declarations, VCS coverage drops to 55.04\%, and technical-limitation coverage drops to 7.16\%. \textit{These results suggest that paper and dataset information are signals of stronger AIBOM documentation readiness}.

\begin{findingbox}
\textbf{Finding 8.} \textit{Models with paper references and dataset declarations provide stronger traceability and limitation coverage than models without these documentation signals.}
\end{findingbox}

\subsubsection{RQ3.2: Task-level documentation coverage}

Table~\ref{tab:task-level-hierarchical-audit} reports task-level AIBOM documentation coverage across paper-reference and dataset branches. The largest task group is \texttt{text-generation}, with 82,591 parsed AIBOM artifacts within the analyzed sample. This task dominates the generated AIBOM dataset, but the hierarchical table also shows coverage differences across multimodal, image-generation, text-classification, and image-classification tasks.

Across tasks, models with both paper references and dataset declarations generally provide stronger coverage for VCS references, technical limitations, safety-risk assessment, and responsible/safety-related fields. For example, in \texttt{text-generation}, models with both paper and dataset information show 99.25\% VCS coverage, 61.59\% technical-limitation coverage, and 66.67\% responsible/safety coverage. In contrast, \texttt{text-generation} models without paper or dataset information show lower coverage, with 49.38\% VCS coverage, 6.66\% technical-limitation coverage, and 9.56\% responsible/safety coverage. Environmental documentation remains nearly absent across task branches.

\begin{table*}
\centering
\caption{Task-Paper-Dataset-level hierarchical audit table of AIBOM documentation coverage.}
\label{tab:task-level-hierarchical-audit}
\scriptsize
\resizebox{\textwidth}{!}{
\begin{tabular}{lllrrrrrrr}
\hline
\textbf{Task} & \textbf{Paper} & \textbf{Dataset} & \textbf{Models} & \textbf{License} & \textbf{VCS} & \textbf{Tech. Lim.} & \textbf{Safety} & \textbf{Resp./Safe.} & \textbf{Env.} \\
\hline
text-generation & Has paper & Has dataset & 4,653 & 2,986 (64.17\%) & 4,618 (99.25\%) & 2,866 (61.59\%) & 1,228 (26.39\%) & 3,102 (66.67\%) & 0 (0.00\%) \\
text-generation & Has paper & No dataset & 2,990 & 2,568 (85.89\%) & 2,952 (98.73\%) & 708 (23.68\%) & 500 (16.72\%) & 1,005 (33.61\%) & 0 (0.00\%) \\
text-generation & No paper & Has dataset & 26,325 & 22,113 (84.00\%) & 10,566 (40.14\%) & 6,213 (23.60\%) & 3,705 (14.07\%) & 7,189 (27.31\%) & 10 (0.04\%) \\
text-generation & No paper & No dataset & 48,623 & 31,437 (64.65\%) & 24,011 (49.38\%) & 3,238 (6.66\%) & 2,338 (4.81\%) & 4,647 (9.56\%) & 1 ($<0.01$\%) \\
\hline
image-text-to-text & Has paper & Has dataset & 412 & 351 (85.19\%) & 412 (100.00\%) & 227 (55.10\%) & 195 (47.33\%) & 275 (66.75\%) & 0 (0.00\%) \\
image-text-to-text & Has paper & No dataset & 389 & 367 (94.34\%) & 389 (100.00\%) & 63 (16.20\%) & 65 (16.71\%) & 90 (23.14\%) & 0 (0.00\%) \\
image-text-to-text & No paper & Has dataset & 603 & 552 (91.54\%) & 603 (100.00\%) & 224 (37.15\%) & 208 (34.49\%) & 253 (41.96\%) & 7 (1.16\%) \\
image-text-to-text & No paper & No dataset & 1,686 & 1,480 (87.78\%) & 1,686 (100.00\%) & 110 (6.52\%) & 192 (11.39\%) & 263 (15.60\%) & 0 (0.00\%) \\
\hline
text-to-image & Has paper & Has dataset & 104 & 86 (82.69\%) & 104 (100.00\%) & 74 (71.15\%) & 56 (53.85\%) & 74 (71.15\%) & 0 (0.00\%) \\
text-to-image & Has paper & No dataset & 104 & 85 (81.73\%) & 104 (100.00\%) & 22 (21.15\%) & 22 (21.15\%) & 24 (23.08\%) & 0 (0.00\%) \\
text-to-image & No paper & Has dataset & 226 & 153 (67.70\%) & 226 (100.00\%) & 32 (14.16\%) & 23 (10.18\%) & 42 (18.58\%) & 0 (0.00\%) \\
text-to-image & No paper & No dataset & 1,214 & 802 (66.06\%) & 1,214 (100.00\%) & 53 (4.37\%) & 20 (1.65\%) & 58 (4.78\%) & 0 (0.00\%) \\
\hline
text-classification & Has paper & Has dataset & 266 & 146 (54.89\%) & 266 (100.00\%) & 147 (55.26\%) & 59 (22.18\%) & 160 (60.15\%) & 0 (0.00\%) \\
text-classification & Has paper & No dataset & 89 & 65 (73.03\%) & 89 (100.00\%) & 6 (6.74\%) & 11 (12.36\%) & 13 (14.61\%) & 0 (0.00\%) \\
text-classification & No paper & Has dataset & 454 & 379 (83.48\%) & 454 (100.00\%) & 172 (37.89\%) & 75 (16.52\%) & 195 (42.95\%) & 0 (0.00\%) \\
text-classification & No paper & No dataset & 457 & 306 (66.96\%) & 457 (100.00\%) & 130 (28.45\%) & 15 (3.28\%) & 140 (30.63\%) & 0 (0.00\%) \\
\hline
image-classification & Has paper & Has dataset & 740 & 717 (96.89\%) & 740 (100.00\%) & 102 (13.78\%) & 30 (4.05\%) & 120 (16.22\%) & 1 (0.14\%) \\
image-classification & Has paper & No dataset & 27 & 27 (100.00\%) & 27 (100.00\%) & 1 (3.70\%) & 0 (0.00\%) & 1 (3.70\%) & 0 (0.00\%) \\
image-classification & No paper & Has dataset & 174 & 152 (87.36\%) & 174 (100.00\%) & 75 (43.10\%) & 10 (5.75\%) & 79 (45.40\%) & 0 (0.00\%) \\
image-classification & No paper & No dataset & 141 & 114 (80.85\%) & 141 (100.00\%) & 36 (25.53\%) & 3 (2.13\%) & 38 (26.95\%) & 0 (0.00\%) \\
\hline
\end{tabular}
}
\end{table*}

\begin{findingbox}
\textbf{Finding 9.} \textit{Task-level AIBOM coverage varies across documentation branches, but paper and dataset availability consistently correspond to stronger traceability, limitation, and responsible-use coverage across model tasks. However, environmental documentation remains weak across all task groups.}
\end{findingbox}

\subsubsection{RQ3.3: Combined license, dataset, and paper-reference coverage}

Table~\ref{tab:aibom-hierarchical-audit} combines license availability, dataset declaration, and paper reference to show how multiple documentation signals shape AIBOM readiness across repository groups. The strongest traceability pattern appears when dataset and paper information are both present. For licensed models with both datasets and paper references, VCS coverage reaches 99.53\%. For unlicensed models with both datasets and paper references, VCS coverage is similarly high at 99.77\%.

However, responsible-use and limitation coverage differ sharply across branches. Models without license, dataset, or paper information show the weakest documentation profile: only 2.87\% include technical limitations, 1.33\% include safety-risk assessment, and 3.66\% include responsible/safety-related information. These results show that AIBOM completeness is not only determined by the generator; the quality of the generated artifact also depends on documentation practices in the source repository.

\begin{table*}
\centering
\caption{Hierarchical audit table of AIBOM documentation coverage across 97,940 generated AIBOM artifacts.}
\label{tab:aibom-hierarchical-audit}
\scriptsize
\begin{tabular}{lllrrrrrr}
\hline
\textbf{License} & \textbf{Dataset} & \textbf{Paper} & \textbf{Models} & \textbf{VCS} & \textbf{Tech. Lim.} & \textbf{Safety} & \textbf{Resp./Safe.} & \textbf{Env.} \\
\hline
Has license & Has dataset & Has paper & 6,373 & 6,343 (99.53\%) & 2,359 (37.02\%) & 1,821 (28.57\%) & 2,742 (43.03\%) & 0 (0.00\%) \\
Has license & Has dataset & No paper & 25,147 & 12,182 (48.44\%) & 5,778 (22.98\%) & 3,841 (15.27\%) & 6,732 (26.77\%) & 16 (0.06\%) \\
Has license & No dataset & Has paper & 4,018 & 3,991 (99.33\%) & 828 (20.61\%) & 600 (14.93\%) & 1,160 (28.87\%) & 0 (0.00\%) \\
Has license & No dataset & No paper & 36,102 & 19,697 (54.56\%) & 3,384 (9.37\%) & 2,437 (6.75\%) & 4,874 (13.50\%) & 1 ($<0.01$\%) \\
\hline
No license & Has dataset & Has paper & 2,216 & 2,211 (99.77\%) & 1,853 (83.62\%) & 148 (6.68\%) & 1,889 (85.24\%) & 1 (0.05\%) \\
No license & Has dataset & No paper & 4,806 & 2,012 (41.86\%) & 1,531 (31.86\%) & 387 (8.05\%) & 1,668 (34.71\%) & 1 (0.02\%) \\
No license & No dataset & Has paper & 621 & 610 (98.23\%) & 124 (19.97\%) & 72 (11.59\%) & 155 (24.96\%) & 0 (0.00\%) \\
No license & No dataset & No paper & 18,667 & 10,450 (55.98\%) & 536 (2.87\%) & 249 (1.33\%) & 683 (3.66\%) & 0 (0.00\%) \\
\hline
\end{tabular}
\end{table*}

\begin{findingbox}
\textbf{Finding 10.} \textit{AIBOM readiness is strongest when repositories combine multiple documentation signals, including license information, dataset declarations, paper references, and VCS links. Repositories lacking these signals provide weaker evidence for AI supply-chain transparency during model reuse.}
\end{findingbox}

\subsubsection{RQ3.4: What types of models appear among the highest-scoring AIBOM artifacts, and how do they differ by origin, architecture, parameter size, and format?}

We selected all artifacts with the highest observed completeness score to examine the upper end of AIBOM completeness. We found that 269 models reach the maximum score of 68.7, and all remain classified as \textit{moderate}. The top-scoring AIBOM artifacts are dominated by large language model families and their redistributed or quantized variants, as shown in Table~\ref{tab:top-aibom-model-categories}. Many top models originate from well-known base families such as Llama, Gemma, Qwen, Granite, Phi, Falcon, CodeLlama, BioMistral, and StableLM. A large share of these repositories use the GGUF format, indicating that quantized or deployment-oriented model variants can still provide enough repository metadata and external references to achieve the highest observed AIBOM score. However, these models remain only moderately complete because metadata and model-card documentation remain limited.

\begin{findingbox}
\textbf{Finding 11.} \textit{Top-scoring AIBOMs are concentrated among well-known LLM families such as Llama, Gemma, Qwen, Granite, Phi, Falcon, CodeLlama, BioMistral, and StableLM. These artifacts achieve strong structural and traceability coverage, but remain only moderately complete because metadata and model-card documentation are still limited.}
\end{findingbox}

\section{Implications}
\label{sec:implications}
We discuss implications for model publishers, software developers, downstream users, AIBOM tool developers, software engineering researchers, and educators.

\textbf{Model publishers.}
Model publishers should treat documentation as part of the released model artifact rather than as optional repository text. Our results show that generated AIBOMs are valid, but AI-specific documentation remains incomplete, especially for model-card fields, limitations, safety-risk assessment, intended use, ethical considerations, environmental information, and meaningful descriptions. More complete model cards, license declarations, dataset references, paper links, and version-control references can directly improve the quality of generated AIBOMs. Publishers should provide structured and machine-readable documentation for model identity, training data, limitations, safety risks, and environmental information. 

\textbf{Software developers, downstream users, and software engineering researchers.}
Developers and downstream users should not treat the existence of an AIBOM as evidence of completeness. Our findings show that many AIBOMs contain required CycloneDX structure and basic model-identification fields, while transparency fields remain missing. Therefore, users should inspect category-level and field-level completeness before relying on AIBOMs for reuse, compliance, auditing, or security review. For software engineering researchers, AIBOM completeness offers an empirical lens for studying AI supply chains, model reuse, documentation evolution, semantic validation, provenance tracking, license compliance, reproducibility, and responsible AI governance.

\textbf{AIBOM tool developers.}
AIBOM tool developers should move beyond artifact generation and provide actionable diagnostics about documentation quality. This need is evident in our results, which show that field presence alone can overestimate transparency when fields contain placeholder values, such as non-informative model descriptions. To address this limitation, future AIBOM tools should distinguish between syntactic completeness and substantive documentation quality. Such tools should report which fields are missing, where missing information should be added, how each missing field affects the completeness score, and which gaps are most relevant for licensing, traceability, safety, environmental reporting, and governance. They could also integrate validation checks for model-card quality, license consistency, dataset traceability, paper-reference availability, and external-resource links.

\textbf{Educators.}
Educators should teach future software engineers and data scientists that using pretrained AI models involves supply-chain, legal, and governance responsibilities. To support this understanding, students should learn how to inspect model cards, licenses, dataset declarations, limitations, external references, and AIBOM artifacts. Courses on software engineering, \gls{ml} engineering, secure software development, and responsible AI should include practical exercises on documenting models, generating AIBOMs, interpreting completeness scores, and identifying missing transparency fields.

\begin{table}[t]
\centering
\caption{Categorization of top-scoring AIBOM models.}
\label{tab:top-aibom-model-categories}
\scriptsize
\begin{tabular}{p{1.55cm}p{1.25cm}p{1.05cm}p{1.05cm}p{1.45cm}}
\hline
\textbf{Origin} & \textbf{Architecture} & \textbf{Size} & \textbf{Format} & \textbf{Examples} \\
\hline
Llama / Meta & Decoder-only LLM & 1B--405B & GGUF & Llama-2, Llama-3.1, Llama-4 \\
Gemma / Google & Decoder-only LLM & 270M--27B & GGUF/HF & Gemma-2, Gemma-3, ShieldGemma \\
Qwen / Alibaba & Decoder-only LLM & 3B--27B & GGUF & Qwen3, Qwen2.5-VL \\
Granite / IBM & Decoder-only LLM & 1B--20B & GGUF & Granite code/base models \\
Phi / Microsoft & Decoder-only LLM & Mini--14B & GGUF/HF & Phi-3.5, Phi-4 \\
Bio/Medical & Domain LLM & 4B--27B & GGUF & BioMistral, MedGemma \\
Code LLMs & Code LLM & 3B--70B & GGUF & CodeLlama, Stable-Code \\
\hline
\end{tabular}
\end{table}

\section{Threats to Validity}
\label{sec:threats}

The threats to validity are organized into three types.

\textbf{Construct validity.}
Our study measures AIBOM completeness using the scoring method implemented by the OWASP GenAI Security Project AIBOM Generator~\cite{owasp_aibom_generator}. This score captures field-level information coverage, but field presence does not always indicate substantive documentation quality and may reflect vague, incomplete, outdated, or non-informative content. To reduce this threat, we complement score-level analysis with granular field analysis and a diagnostic check for meaningful descriptions. Therefore, our results should be interpreted as evidence of AIBOM documentation completeness, not as a full assessment of model quality, safety, fairness, or performance.

\textbf{Internal validity.}
Our pipeline depends on the \gls{hf} API, the OWASP AIBOM Generator, and our parsing scripts. API changes, unavailable repositories, rate limits, tool failures, or inconsistent metadata could affect artifacts and extracted scores. To reduce this threat, we preserved missing values, stored JSON artifacts locally, linked each score record to its artifact, removed duplicate identifiers, and excluded artifacts that failed parsing or lacked score information. However, limitations in the generator's extraction logic may influence which fields appear in AIBOMs. We also inspected 100 AIBOM artifacts and score reports to check the consistency of the pipeline.

\textbf{External validity.}
Our dataset is based on a snapshot of 2,942,466 \gls{hf} repositories and focuses on models with more than 100 downloads. This filter supports analysis of models with observable reuse, but findings may not generalize to low-download models, private repositories, gated models, enterprise registries, or models hosted outside \gls{hf}. The \gls{hf} ecosystem changes rapidly as model-card templates, repository practices, and AIBOM tools evolve. Therefore, completeness patterns may change in later snapshots or model-hosting ecosystems.

\section{Conclusion and Future Work}
\label{sec:conclusion}

This paper studied the completeness of AI Bills of Materials generated from pretrained machine learning models hosted on the \gls{hf} Hub. We analyzed approximately 97.5K generated AIBOM artifacts and examined completeness at the score, category, granular-field, and repository-characteristic levels for AI supply-chain transparency. Results indicate that generated AIBOMs are structurally valid but only moderately complete overall. Required CycloneDX fields are consistently present, and component-basic information is comparatively well represented. However, metadata and model-card documentation remain limited, with model-card documentation showing the weakest coverage. At the field level, important AI-specific transparency information is often missing or weakly represented, including intended use, ethical considerations, safety-risk assessment, environmental information, technical limitations, and meaningful descriptions. In future work, we aim to analyze further dimensions of AIBOM quality beyond field presence, including the semantic quality of descriptions, limitations, intended-use statements, and safety-risk assessments.

\section*{Declaration of Generative AI Technologies }
During manuscript preparation, the authors used ChatGPT only to improve flow, grammar, and clarity. The tool was not used to generate technical content, synthesize citations, or verify experimental facts. The authors reviewed and verified all outputs and take full responsibility for the final manuscript.

\bibliographystyle{ACM-Reference-Format}
\bibliography{sample-base}

\end{document}